%% file: ms.tex
\documentclass[apj, numberedappendix]{emulateapj}
\usepackage{xspace}
\usepackage{color}
\usepackage{hyperref}
\usepackage{afterpage}
\hypersetup{
   pdftitle={Precision Measurement of The Most Distant Spectroscopically Confirmed Supernova Ia with the Hubble Space Telescope}, 
   pdfauthor={David Rubin},
   colorlinks=true,        
   linkcolor=blue,         
   citecolor=blue,         
   urlcolor=blue           
}

\newcommand{\ang}{\AA\xspace}

\newcommand{\mstrue}{m^{\mathrm{true}}_{\star}}

\newcommand{\loms}{\mstrue < 10^{10} M_{\sun}}

\slugcomment{ApJ Volume 763 Number 1}
\shorttitle{Redshift \twodecimalz Supernova}
\shortauthors{Rubin et al.}
\input{numbers.tex}

\begin{document}

\title{Precision Measurement of The Most Distant Spectroscopically Confirmed Supernova I\lowercase{a} with the \textit{Hubble Space Telescope}\footnotemark[1]}

\footnotetext[1]{Based on observations with the NASA/ESA \textit{Hubble Space Telescope}, obtained at the Space Telescope Science Institute, which is operated by AURA, Inc., under NASA contract NAS 5-26555, under programs GO-9583, GO-9727, GO-9728, GO-10339, and, GO-11600.}

\input{auth.tex}

\begin{abstract}
We report the discovery of a redshift \twodecimalz supernova in the GOODS North field. The \textit{Hubble Space Telescope} (\textit{HST}) ACS spectrum has almost negligible contamination from the host or neighboring galaxies. Although the rest frame sampled range is too blue to include any Si {\sc ii} line, a principal component analysis allows us to confirm it as a Type Ia supernova with 92\% confidence. A recent serendipitous archival \textit{HST} WFC3 grism spectrum contributed a key element of the confirmation by giving a host-galaxy redshift of \mbox{$\galaxyz \pm \galaxyzerr$}. In addition to being the most distant SN Ia with spectroscopic confirmation, this is the most distant Ia with a precision color measurement. We present the ACS~WFC and NICMOS~2 photometry and ACS and WFC3 spectroscopy. Our derived supernova distance is in agreement with the prediction of $\Lambda$CDM.
\end{abstract}


\section{Introduction}

Over the past 15 years, \textit{HST} has played an integral role in measuring cosmological parameters through the Type Ia supernova Hubble diagram \citep{perlmutter97, garnavich98, riess98, perlmutter99, knop03, riess04, riess07, amanullah10, suzuki12}. With its low background and diffraction-limited imaging, \textit{HST} is capable of measuring supernovae at redshifts that are very difficult from the ground. Measuring very distant supernovae breaks degeneracies in the lower-redshift Hubble diagram, enabling us to probe the nature of dark energy at redshifts above $z\sim0.5$ independently of its low-redshift behavior. In this paper, we present the most distant cosmologically useful supernova to date and show that even at this distance, \textit{HST} can still make measurements with precision.

\section{Search and Followup}

\scpname was found in the GOODS North Field \citep{dickison03} as part of a supernova survey with sets of supernova followup that were alternated between the Supernova Cosmology Project (SCP)\footnote{HST GO Program 9727} and the Higher-Z SN Search Team\footnote{HST GO Program 9728}. Four epochs of Advanced Camera for Surveys (ACS) F850LP and F775W (these are $z$ and $i$-band filters) observations were obtained, with a cadence of $\sim7$ weeks. In the first cadenced epoch (2004 April 3), this candidate was discovered in the reference-subtracted\footnote{The reference images for this field come from Program ID 9583.} F850LP image with a signal-to-noise ratio of \snStoNz (Vega magnitude \snfoundmagz, see details of photometry in Section \ref{sec:photometry}). In the concurrent F775W image, it had a signal-to-noise ratio of \snStoNi (Vega \snfoundmagi). Because the red observed color implied a possible very-high-redshift SN Ia, we followed it with ACS F850LP and Near Infrared Camera and Multi-Object Spectrometer (NICMOS~2) F110W and F160W (very broad $J$ and $H$-band filters) photometry, and ACS G800L grism spectroscopy\footnote{This supernova is referred to in the HST archive as SN150G and elsewhere by its nickname ``Mingus'' \citep{gibbons04}.}.

\begin{figure}[h]
   \centering
   \includegraphics[width=3.4in,  clip=True]{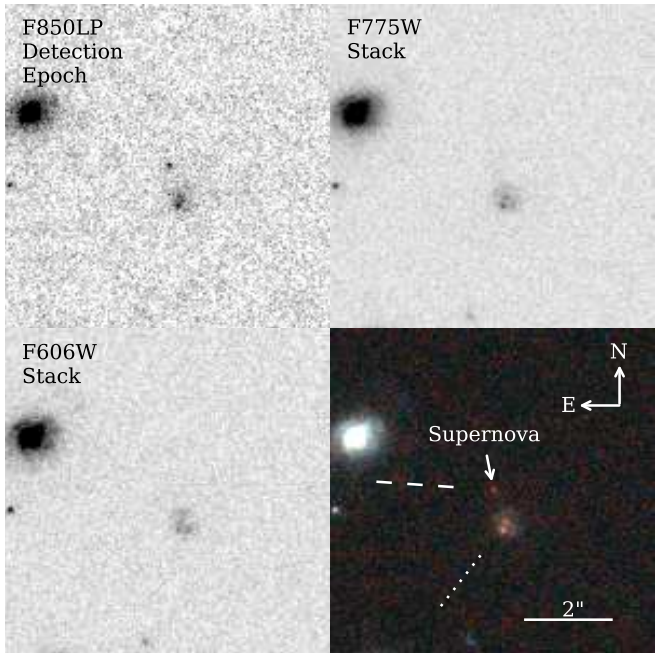} 
   \caption{ACS images of the supernova location.  The lower right panel shows a
three-component color image composed from: an F606W stack (blue), F775W
stack (green), and the F850LP SN detection epoch (red), which are shown
in the remaining panels. The lines indicate the dispersion direction in ACS (dashed) and WFC3 (dotted) spectroscopy. The supernova coordinates are 12:37:09.5 +62:22:15.5 (J2000.0).}
   \label{fig:acsimage}
\end{figure}

The sky in the vicinity of the SN is shown in Figure \ref{fig:acsimage}\footnote{In addition to the other datasets, data from HST GO Program 10339 was used for this figure and the subsequent host-galaxy analysis.}. The likely host is the late-type galaxy at redshift \galaxyz (see Section \ref{sec:hostspec}) centered \sntocore away. This corresponds to only \sntocorekpc if the SN and galaxy are at the same distance. Light from this galaxy is visible at the location of the supernova and no other galaxies down to a magnitude limit of \nogallimit are within \nextclosest. In the F775W and redder data, this galaxy has two cores, indicating a possible merger. The consistency of the colors of these cores (always $< 0.3$ mag, typically $< 0.1$) over the wide range of 4350\ang to 16000\ang makes it extremely likely that these cores are at the same redshift.

\section{Spectroscopy}

\subsection{ACS Grism Observations of SN and Host}\label{sec:acsgrism}
We obtained eleven orbits of spectroscopy with the ACS G800L grism nine days after the discovery epoch. The light curve fit (Section \ref{sec:lcfit}) indicates that the spectrum was taken \specphase rest-frame days after rest-frame $B$-maximum. We extracted spectra for the likely host and SN with aXe \citep{kummel09}. No conclusive features or lines were apparent in the spectrum of the galaxy, nor did the two cores give significantly different spectra.

\subsection{Wide Field Camera 3 Grism Observations of the Host}\label{sec:hostspec}
As a fortunate coincidence, two orbits of Wide Field Camera 3 (WFC3) IR G141 grism spectroscopy were taken in this region of GOODS North on 2010-09-26\footnote{Data from HST GO Program 11600}. Although the F140W direct image missed the host galaxy, the grism dispersed the host into the field of view. Matching objects between ACS F850LP imaging and the direct image allowed us to compute the position of the host galaxy for use by aXe.

\begin{figure}[h]
   \centering
   \includegraphics[width=3.5in,  clip=True]{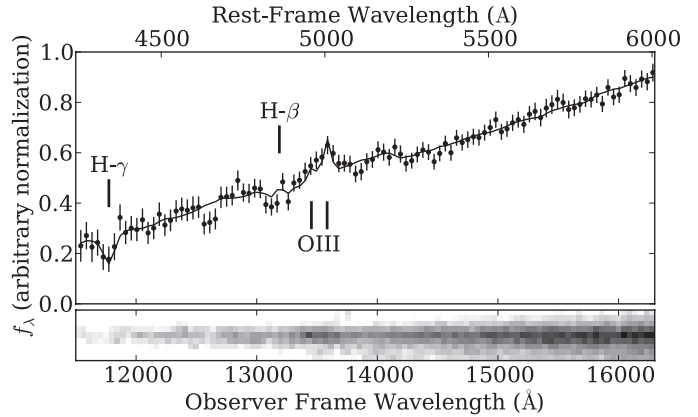} 
   \caption{Upper panel: Extracted WFC3 IR spectrum of the likely host galaxy with template fit using SDSS galaxy principal components (solid line). The best-fit (and only reasonable) redshift is \galaxyz. We note that including the ACS grism data for the host (5500\ang to 10000\ang) has no effect on the fit. Lower panel: 2D WFC3 spectrum, spanning 103 pixels. Some of the flux visible at longer wavelengths than the features is contamination.
   \label{fig:galspec}}
\end{figure}

The host galaxy spectrum is shown in Figure \ref{fig:galspec}, along with the best-fit template derived by scaling principal components of SDSS spectra \citep{aihara11}. Only one feature is detected at very high statistical significance: an emission feature at $13600$\ang. The only reasonable match to the spectrum between redshift 1.0 and 2.0 is one centered on redshift \galaxyz. The emission feature is then made up of a blend of the [O{\sc iii}]$\lambda \lambda$ 4959, 5007\ang doublet. No other emission lines are required to appear in the wavelength range of either grism spectrum for this to be a credible template match. We also see possible absorption from H$\gamma$ and H$\beta$ (4340\ang, and 4861\ang rest-frame wavelengths, respectively), but at lower statistical significance. As we are not sure which core (or both) emit the [O{\sc iii}], we take a conservative $0.1\arcsec$ separation $= 36$\ang systematic uncertainty in the observer-frame wavelength of the lines. This translates to a \galaxyzerr uncertainty on the redshift, which dominates the other sources of uncertainty.

\begin{figure*}[!h]
   \centering
   \includegraphics[width=3.5in,  clip=True]{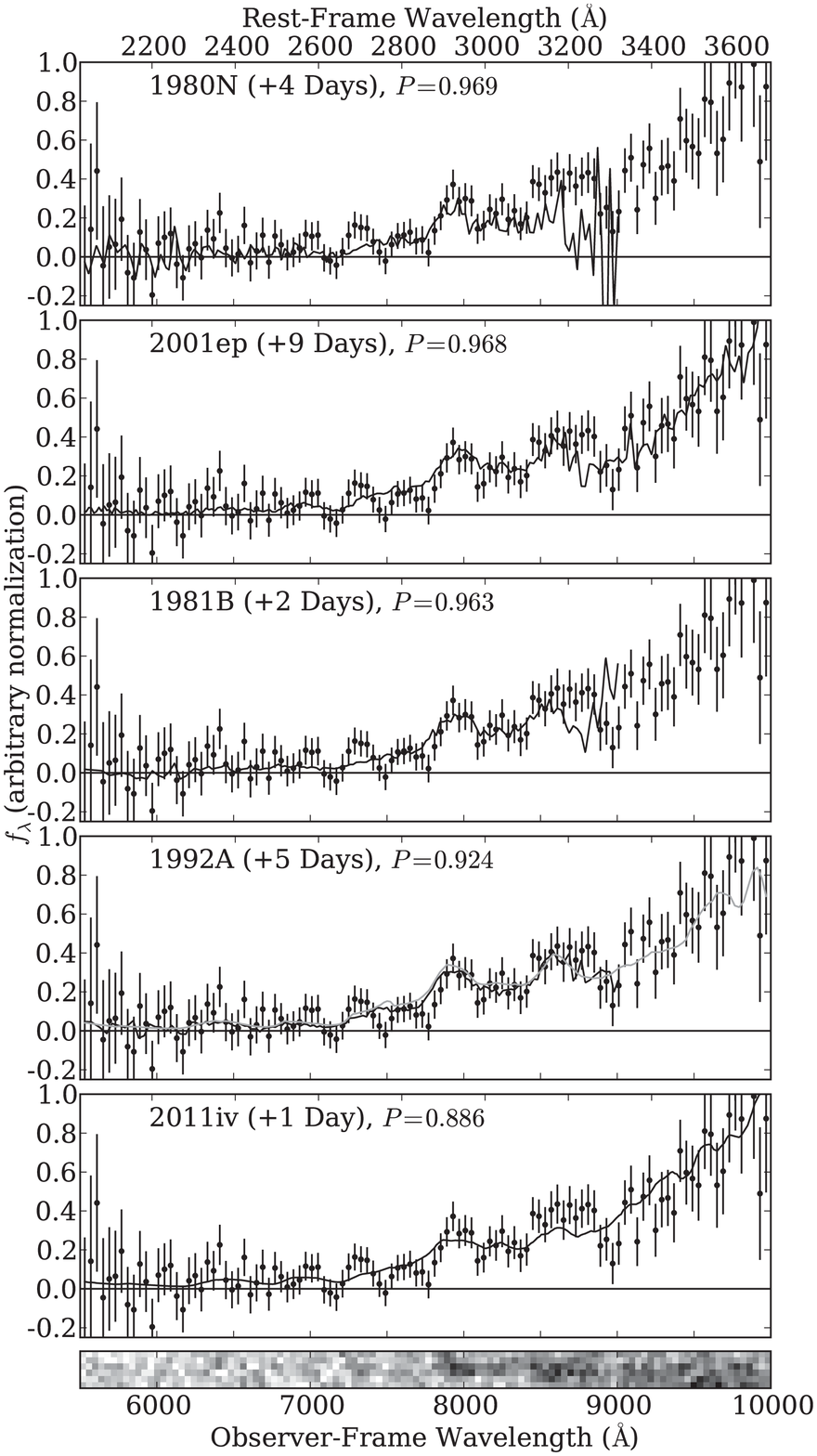}
   \hspace{-0.1 in}
   \includegraphics[width=3.5in,  clip=True]{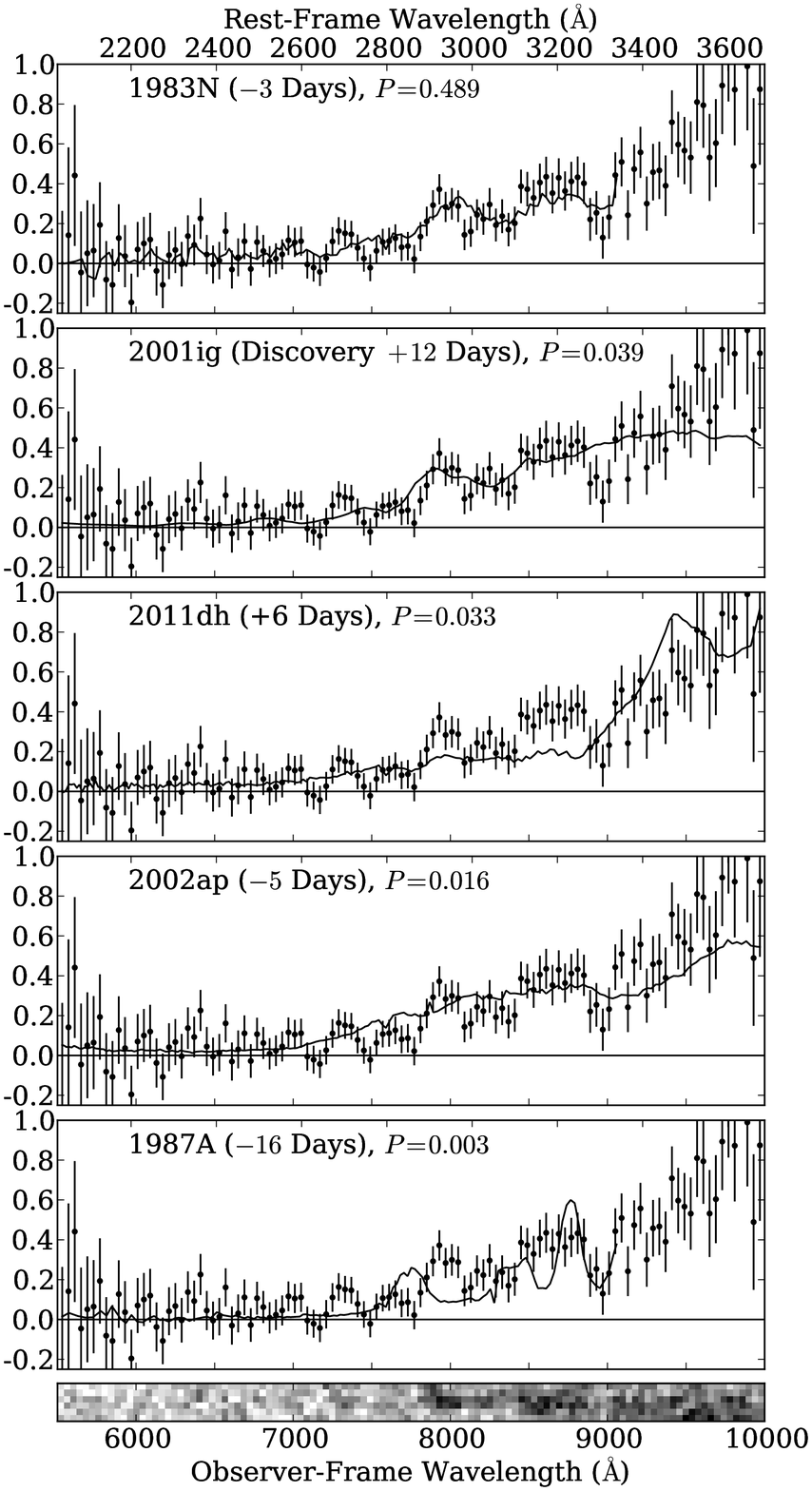} 
   \caption{Each panel shows a comparison between \scpname (points with error bars) and another SN. The five best-matching comparison SNe Ia are shown in the left panels; the five best-matching comparison CC SNe are shown in the right panels. For each comparison SN, only the best-matching epoch is shown. The best visual match is SN1992A (left, third from top); we have overlaid additional data from a phase of +8 days that covers the full rest-frame wavelength range (light grey), showing that the match continues for the full spectrum. Of the 17 CC SNe (the best five of which are shown here), only SN1983N is a possible match, although as noted in the text, this SN is two magnitudes fainter at max than a typical SN Ia. \\
Bottom panels: 2D \scpname spectrum, spanning 112 pixels. Some of the flux visible in the very reddest wavelengths is contamination from a nearby galaxy.
   \label{fig:snspec}}
\end{figure*}

\subsection{Typing}\label{sec:typing}

aXe resamples the grism data, correlating neighboring flux measurements. This can be seen by eye in the spectrum (points with errorbars in Figure \ref{fig:snspec}), that is, the difference between neighboring flux measurements is generally smaller than one would expect from the indicated error bars. These positive correlations reduce the statistical significance of spectral features, so a quantitative understanding of these correlations is crucial. By examining blank sky, we find that the correlation between neighboring errors is 0.4 (and confirm the accuracy of the on-diagonal errors reported by aXe). The weight of the spectrum scales with the correlation between neighbors ($\rho$) as  $1/(1 + 2\rho)$ (see Appendix \ref{sec:matinverse} for the derivation). The weight of the spectrum is thus reduced by 44\% compared to a naive reading of the aXe error bars. All $\chi^2$ values in this paper are computed using a covariance matrix containing nearest-neighbor correlations.

As our supernova spectrum misses the Si {\sc ii} $\lambda 6355$ \citep{wheeler85, uomoto85, panagia85} and Si {\sc ii} $\lambda4130$ \citep{clocchiatti00} lines normally used for confirming SNe Ia, we use statistical methods for classifying \scpname.

We first begin by collecting the comparison rest-frame UV spectra available to us. A useful list of SNe observed with the \textit{HST} and the \textit{International Ultraviolet Explorer} (\textit{IUE}) is \citet{panagia03}, with an updated list, including \textit{Swift}-observed, in \citet{brown09}. We obtained \textit{IUE} spectra and \textit{HST} spectra from the Mikulski Archive for Space Telescopes (MAST)\footnote{\url{http://archive.stsci.edu/}}, and Swift spectra from the SUSPECT archive\footnote{\url{http://suspect.nhn.ou.edu}}. More recent spectra were found by searching MAST, others came from the literature. A summary of all data is given in Table \ref{tab:spectra}; we collect 94 spectra in total from 33 SNe, all within $\sim 2$ weeks of $B$ or $V$-maximum (whichever is quoted in the literature).

\input{uvtable}

Our goal is to compare these spectra to the spectrum of \scpname, extracting a probability of matching for each. Unfortunately, most of the data are from the IUE, and only extend to $\sim 3300$\ang observer-frame, rather than 3600\ang as we have with \scpname (a related issue is the presence of noise in the comparison spectra). This limitation complicates the comparison of these spectra to \scpname. 

Another, more subtle, issue is also relevant. We note that simply converting a $\chi^2$ per degree of freedom to a probability \citep[e.g., ][]{rodney12} is never appropriate when comparing different models to the same data. $\Delta \chi^2$ values (the difference in $\chi^2$ between models) can be converted into probabilities, but this requires knowing the dimensionality of the parameter space\footnote{A well-known example is the $68.3\%$ confidence interval, which is given (in the assumption of Gaussian errors and an approximately linear model) by $\Delta \chi^2 < 1$ in one dimension and $\Delta \chi^2 < 2.30$ in two.}.

We can address both issues (limited coverage and estimating dimensionality) by performing a principal component analysis of all spectra in the UV. The details are discussed in Appendix \ref{sec:spectralmodel}. After computing the mean and first two principal components, we can compute a $\Delta \chi^2$ between \scpname and every other spectrum in turn. We fit \scpname and another spectrum with the projections onto the components constrained to be the same (we allow them to have different normalizations); this gives us a joint $\chi^2$. We then subtract the $\chi^2$ values for \scpname and the other spectrum when they are allowed to have different projections. This $\Delta \chi^2$ value gives us the probability that the SNe have different true projections given the observed data. We then subtract this value from 1 to get a ``matching probability.''

These results are summarized in Table \ref{tab:prob}. Thirteen SNe have matching probabilities above 0.05; twelve of these (and all of the top six) are SNe Ia. The average matching probability of a SN Ia is 41.8\%; the average probability for a core-collapse SN is 3.4\%. The probability of \scpname being a Ia from the spectrum alone (assuming an equal fraction of SNe Ia and CC SNe; see below) is therefore $41.8/(3.4 + 41.8) = 92\%$. In Figure \ref{fig:snspec}, we plot the best-matching spectrum of the five best-matching SNe of each type. Of the CC SNe, only SN1983N is a credible match spectroscopically, although this supernova was two magnitudes fainter at maximum than a typical SN Ia \citep{1985BASI...13...68P}.

\input{probabilities}

\afterpage{\clearpage}

We now must evaluate the relative ratio of CC SNe to SNe Ia at redshift \galaxyz for SNe with comparable brightness to SNe Ia. \citet{bazin09} present both photometrically and spectroscopically classified SNe from the Supernova Legacy Survey and the associated absolute magnitudes (their definition is similar to a $V$-band AB absolute magnitude). For SNe with brightness comparable to most SNe Ia ($\sim -19$), they find a SN Ia to CC rate of $\sim$5-to-1 at redshift 0.3. However, at redshift \galaxyz, the star-formation rate is $\sim 5$ times higher than at redshift 0.3 \citep{hopkins06}, raising the core-collapse rate by approximately the same value. The SN Ia rate is equal to within the error bars (tens of percent) at redshift 0.3 and redshift \galaxyz \citep{barbary12}, so both classes of SNe are comparably common at this redshift. We therefore retain the $92\%$ confidence that was derived ignoring the rates.

It is also encouraging that the spectrum of \scpname matches the theoretical SN Ia spectra of \citet{lentz00} derived from the W7 model \citep{nomoto84} (see Table \ref{tab:prob}). The best match is for the unscaled heavy element abundance (that is, no change from W7).

As a less-likely possibility, we investigate the possibility that the nearby galaxy is not the host. We use the spectra with broad wavelength coverage (almost all of those in Table \ref{tab:spectra} except the IUE spectra) and match them against \scpname with the redshift floating. It is reassuring that the best match is a Ia (SN1992A) at redshift 1.72, at least for this limited set of SNe.

This analysis may turn out to be conservative. In the Lick Observatory Supernova Search volume-limited sample \citep{Li11}, the ratio of SNe II to SNe Ibc is about 3-to-1, similar to what we have in our sample of spectra. However, the SNe Ibc are fainter on average than SNe II; in \citet{bazin09}, the ratio appears to be higher (in the luminosity range of SNe Ia). If SNe Ibc are the only plausible non-Ia match to \scpname, then our confidence that \scpname is a SN Ia may get stronger simply from revised rates. It is also possible that no SNe Ibc are credible matches to \scpname, and more wavelength coverage of SN1983N would have shown us that it does not match. In the future, additional core-collapse comparison spectra will resolve this question.

\section{SN Photometry}\label{sec:photometry}
We used similar techniques for the SN photometry as were used in \citet{suzuki12}; these are summarized below. In the spirit of ``blinded'' analysis, we finalize the photometry before looking at the light curve or distance modulus. We give our photometry in Table \ref{tab:phot}.

\input{photometry}

\subsection{ACS Photometry}
We begin by iteratively combining each epoch with MultiDrizzle \citep{fruchter02, koekemoer02} and aligning all epochs. Aperture photometry with a three-pixel radius ($0.15\arcsec$) is computed for all epochs, with the zero level set by the many epochs without the SN. As the pixel values in the resampled images are correlated, the background error is derived empirically (by placing many three-pixel radius apertures in object-free parts of the image), and the Poisson error of the aperture flux is added in quadrature. We use a zeropoint of \zZP for the F850LP data, derived in \citet{suzuki12} along with the effective throughput, and \iZP for the F775W data, from \citet{bohlin07}.

\subsection{NICMOS Photometry}
The optimal radius for aperture photometry with NICMOS is approximately 1 pixel ($0.076\arcsec$), precluding any resampling of the NICMOS images. Following \citet{suzuki12}, we therefore performed the NICMOS photometry using analytic galaxy models (one for each filter) which were convolved with their PSFs and resampled to match the images. The supernova position and fluxes were modeled simultaneously using PSFs generated for each spectral energy distribution (SED) and band. As there are two cores for this galaxy, we use two azimuthally symmetric elliptical models (with radial variation described by splines) to model the cores (as the SN is reasonably far off-core, this is mainly needed to get the centroid of the model correct for each image). The remaining azimuthal asymmetry of the galaxy was modeled with a two-dimensional second-order spline, with nodes spaced every five pixels ($0.38\arcsec$).

While optimizing the host-galaxy model (e.g.,~the spline-node spacing), we use simulated SNe at dozens of positions at comparable separation from the galaxy to check for any bias or unexplained variance in the photometry. No bias is seen at the 0.01 magnitude level in either band. However, the final epoch in F110W shows a small amount of unexplained variance ($\chi^2$/degree of freedom 1.35) for the recovered fluxes around the true flux, possibly due to slight South Atlantic Anomaly persistence. We rescale the photometry error bar for this epoch to make the $\chi^2$ per degree of freedom 1.

We used a NICMOS F110W zeropoint of \nicJzp \citep{ripoche12} and a NICMOS F160W zeropoint of \nicHzp \citep[see discussion in][]{amanullah10}.

\section{Analysis}
\subsection{Light-Curve Fit}\label{sec:lcfit}

\begin{figure}[t]
\centering   
   \includegraphics[width=3.5in, clip=True]{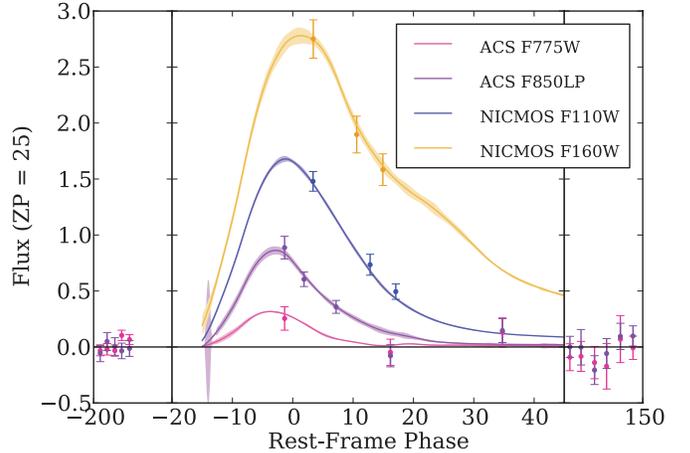} 
   \caption{SALT2-2 fit to the photometry. To illustrate the quality of the F775W data, the F775W photometry is shown in this plot; as it is too blue for SALT2-2 to fit reliably, these data are not used in any analysis. The error snakes represent the model errors of SALT2-2.}
      \label{fig:lcplot}
\end{figure}

We fit the light curve of the SN with SALT2-2 \citep{guy10}, a principal component expansion of type Ia supernova SEDs. The fit parameters are the date of rest-frame $B$-band maximum, the magnitude at maximum ($m_B$), the broadband color ($c$, similar to a rest-frame $B-V$ color), and light-curve width ($x_1$, the next supernova principal component after color). We find $m_B, x_1, c = (26.14, 0.2, -0.10).$ The best-fit template is shown in Figure \ref{fig:lcplot}. The corrected-distance-modulus statistical error is only \cormagerr mag. (This value does not include Hubble diagram dispersion that cannot be removed with the magnitude corrections detailed in Section \ref{sec:conclusions}.) As we lack a measurement on the rise of \scpname, the date-of-maximum constraints are asymmetric. We derive the distance modulus uncertainty by sampling from the true corrected distance modulus distribution (by running a Metropolis-Hastings Monte Carlo using the SALT2-2 model). There is a fortuitous cancellation between the date of maximum and the light curve parameters: moving the date of maximum earlier brightens the peak magnitude while increasing the light-curve width and making the color slightly bluer. After applying the corrections in Section \ref{sec:conclusions}, the corrected-magnitude likelihood is well-constrained (and is Gaussian).

\subsection{Host Stellar Mass}
As SALT2 Hubble residuals are correlated with host-galaxy stellar mass \citep{kelly10, sullivan10}, we must estimate the host mass for \scpname. We used a Z-PEG \citep{leborgne02} fit to broad-band galaxy photometry, similar to the methods used in those papers. Using aperture photometry with a $1\arcsec$ radius, and zeropoints from \citet{bohlin07}, we derived the following AB magnitudes for the host galaxy: 25.7 (F435W), 25.2 (F606W), 24.2 (F775W), 23.4 (F850LP), and 20.0 (F160W, Vega = 0). To accurately fit all photometry, Z-PEG requires a template with age 5 Gyr, which is older than the universe at this redshift (4 Gyr). The stellar mass confidence interval when enforcing an age-of-the-universe constraint is essentially contained inside the confidence interval when allowing age to be unconstrained. To be conservative, we do not enforce this constraint, obtaining a $\log_{10}$ stellar mass of $11.2^{+0.1}_{-0.4}$, easily putting this galaxy inside the high-mass ($>10^{10} M_{\sun}$) category.

\subsection{Systematic Errors}
\subsubsection{Calibration}
Fitting an accurate corrected magnitude requires fitting an accurate color ($c$). The farther apart the filters used are in wavelength, the less (uncorrelated) calibration uncertainties affect the derived $c$, and therefore the derived corrected magnitude. For a given range of wavelength coverage, measuring a supernova in more filters will also decrease the sensitivity of the fit to any given miscalibration (again assuming independent calibration uncertainties for the data in each filter). With three passbands within the SALT2-2 range and a long wavelength baseline, the SN distance modulus we derive from the light curve fit is more resilient against calibration uncertainties than most high-redshift SNe distances. Our distance modulus is most sensitive to the F160W zeropoint, with $\partial \mu/\partial $(F160W zeropoint) =1.5 (that is, a change in the F160W zeropoint of 0.01 magnitudes changes the corrected magnitude by 0.015), a factor of two better than is typically achieved with only one color. The other calibration uncertainties combine to a systematic error of only $\sim 0.01$ mag on the distance modulus.

The NICMOS 2 F160W data are affected by a count-rate nonlinearity of \nicHCRNLerr \citep{dejong06}, which adds an effective zeropoint uncertainty of 0.06 magnitudes at the flux level of high-redshift SNe, assuming a power-law dependence of the non-linearity over the full range of flux between the standard stars and the SNe (4-5 dex). Based on the F110W results of \citet{ripoche12}, we add an uncertainty of 0.03 magnitudes to account for possible deviation from a power law. We will improve this uncertainty with a future recalibration of the F160W non-linearity using the techniques in Ripoche et al. in a future paper.

\subsubsection{Malmquist Bias}
Most SNe Ia at redshift \twodecimalz would be too faint to be found by the search, even at maximum brightness. Malmquist bias is therefore present. Most of this bias is taken out by making the corrections we describe in Section \ref{sec:conclusions}, but some bias remains. (If it were possible to perfectly correct SNe, such that all SNe were equally bright after correction, no Malmquist bias would remain.) A simple simulation (detailed further in Rubin et al., in prep) that selects SNe from the parent distribution and determines if they could be found at redshift \twodecimalz allows us to estimate that this remaining Malmquist bias is about 0.08 mag.

If there are SNe at high enough significance to find, but not to get a spectrum of, there may be additional Malmquist bias. We investigate this possibility here using the observed spectrum of \scpname. The faintest supernova we could have found would be S/N $\sim5$, rather than S/N $\sim9$. Increasing the noise in the spectrum by a factor 1.8 allows more supernovae of both types to match the spectrum. The net effect is to lower the confidence of being a Ia to 86\%, in which case we would still use the supernova for cosmological analysis. (In an earlier study \citep{kowalski08}, we showed that the analysis is robust to this level of non Ia contamination.)

The largest contributors to the Malmquist bias uncertainty are the magnitude cut for the search (which we take to be uncertain at the 0.2 mag level) and the uncorrected residual dispersion of SNe at redshift \twodecimalz (which we take to be $0.20\pm0.05$ (see discussions below in Sections \ref{sec:lensing} and \ref{sec:conclusions}). Each of these contributes about 0.03 magnitudes to the Malmquist bias uncertainty. Therefore, the total uncertainty, which would correlate from supernova-to-supernova were there others like it, is about 0.04 mag.

\subsubsection{Lensing}\label{sec:lensing}

The bright spiral galaxy $3.5\arcsec$ away from the supernova (visible to its upper left in Figure \ref{fig:acsimage}) is at redshift 0.64 \citep{cowie04}, and is thus a potential source of gravitational magnification for the supernova. Here, we provide a rough estimate of the size of this effect.

As with the host galaxy, we used Z-PEG to derive the stellar mass. For this larger (apparent size) galaxy, we used a $1.5\arcsec$ radius, and obtained the following AB magnitudes: 23.5 (F435W), 22.7 (F606W), 21.8 (F775W), and 21.5 (F850LP). We use the Z-PEG stellar mass of \spiralstellarmass with the relation between stellar mass and halo mass from \citet{guo10} to derive the total mass of the halo, \spiralhalomass. Assuming a singular isothermal sphere model, with $M_{200} \sim M_{\mathrm{halo}}$, we find a magnification of 1.08 (using the \citet{nfw96} NFW profile provides virtually the same answer). This number is not the magnification of the supernova; had the lensing galaxy not been there, the supernova would likely be slightly de-magnified (compared to a filled-beam distance modulus). \citet{holz05} find that the scatter due to lensing is approximately $0.093 z = 0.16$ magnitudes at this redshift. We include this uncertainty in our distance modulus error (as noted below) and see no evidence that \scpname is magnified or de-magnified by more than this.

The mean magnification of supernova fluxes is zero at a given redshift. (Selection effects can bias the observed SNe to higher magnification, but \citet{jonsson06} see no evidence of this in the \citet{riess04} sample.) However, we fit our cosmological constraints in log(flux) (magnitudes), where the mean magnification is not zero (as supernova fluxes are roughly log-normally distributed, and we use least-squares fitting, fitting in magnitudes is appropriate). We evaluate the lensing bias from working with magnitudes using the distributions of \citet{wang02} and find it to be 0.01 mag (biased faint). In principle, most of this bias is well-understood (from knowledge of the corrected supernova luminosity distribution and the lensing distribution) and could be removed.

\section{Conclusions}\label{sec:conclusions}
   
 \begin{figure*}[ht]
   \centering
   \includegraphics[width=4.5in, clip=True]{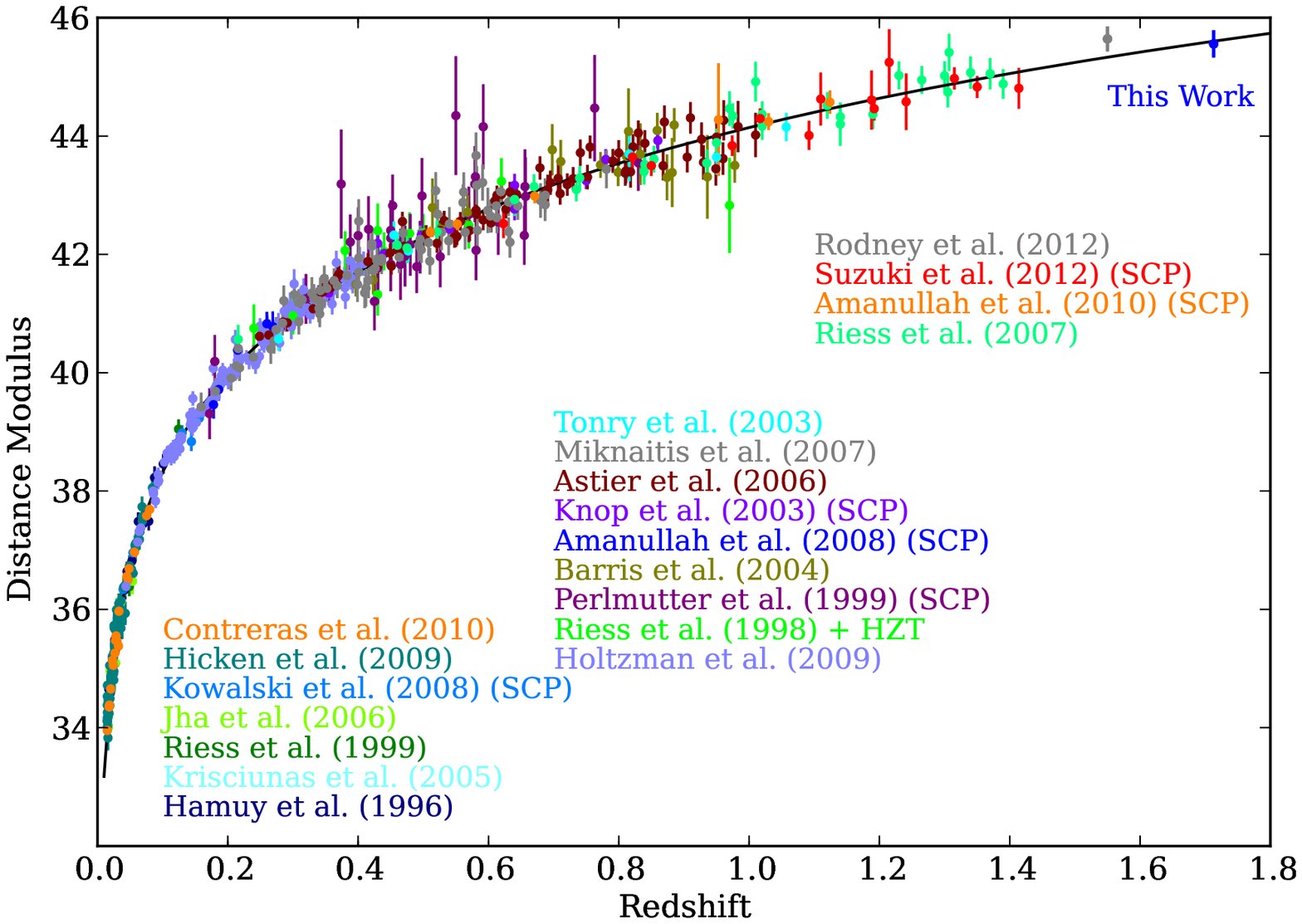} \\
      \includegraphics[width=4.5in, clip=True]{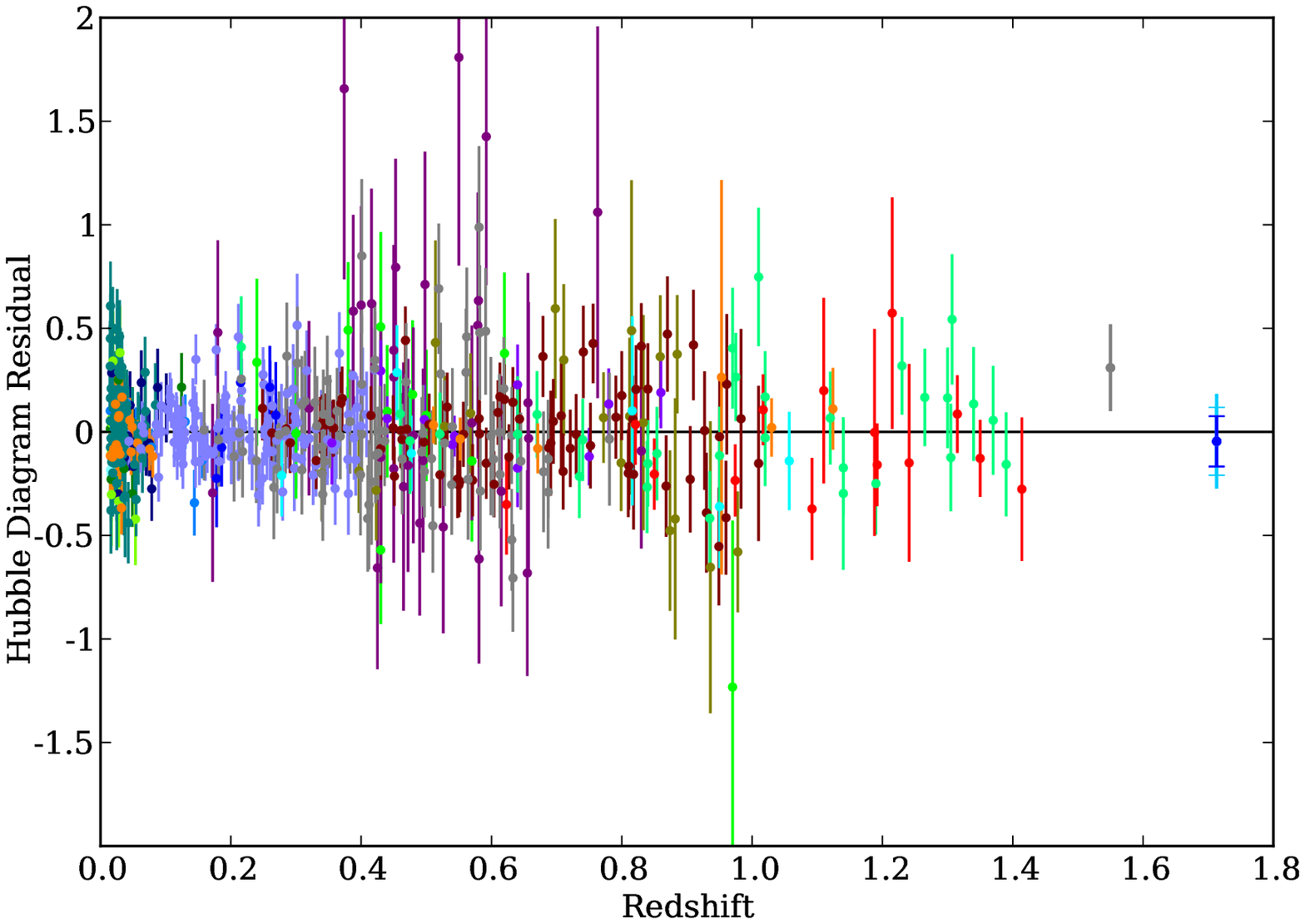} 
\caption{ Top Panel: \citet{suzuki12} Hubble diagram (with the best-fit flat $\Lambda$CDM model) with Primo \citep{rodney12} and \scpname added.\\
 Bottom Panel: Hubble diagram residuals. The inner (blue) error bars on \scpname show the uncertainty of the light-curve fit. The middle (capped, cyan) error bars include the sample dispersion; the outer error bars include the lensing dispersion. Future analyses including spectral information or gravitational lensing correction might improve these outer error bars.}
   \label{fig:hubblediagram}
\end{figure*}

\begin{figure*}[ht]
   \centering
   \includegraphics[width=7in, clip=True]{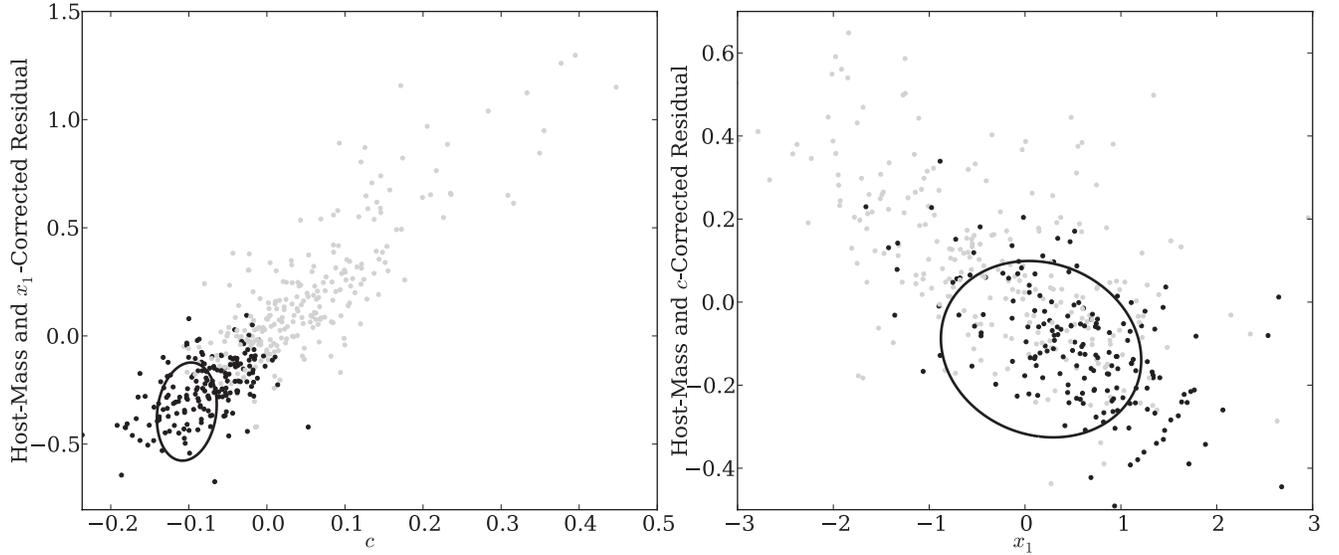} 
   \caption{Plot of Hubble residuals (from the best-fit flat $\Lambda$CDM model) against $c$ (left panel) and $x_1$ (right panel). In the left panel, the distance moduli have been corrected for $x_1$ and host mass, revealing the $c$-brightness relation. Similarly, the distance moduli in the right panel are corrected for $c$ and host mass. Each ellipse represents the ($\Delta \chi^2 = 1$) SALT2-2 Gaussian approximation to the likelihood for \scpname; projecting the uncertainty to each axis gives the 1-$\sigma$ error bars on each parameter. The points are comparison supernovae taken from Rubin et al. (in prep); for clarity, only SNe measured to better than 0.05 mag in $c$ are shown. The black points represent SNe that would be bright enough in F850LP (at peak) to have been found at redshift \twodecimalz in our search.   \vspace{0.2in}}
   \label{fig:alphabeta}
\end{figure*}

We apply the corrections detailed in \citet{suzuki12} Equation 3 to obtain a SALT2-2 distance modulus corrected for $x_1$, $c$, and host mass, reproduced here.
\begin{equation}
  \mu_B = m_B + \alpha\cdot
  x_1 - \beta\cdot c + \delta \cdot P(\loms) - M_B\, ,\label{eq:magcor}
\end{equation}
 where $\alpha$ is the light-curve-width-correction coefficient, $\beta$ is the color-correction coefficient, $\delta$ is the host-mass-correction coefficient, and $M_B$ is the ($h = 0.70$) absolute $B$-magnitude. In addition to the propagated lightcurve fit uncertainties, we add in quadrature the distance modulus scatter due to lensing (above) and $\sigma_{\mathrm{sample}}$, the error needed to get a $\chi^2$ per degree of freedom of 1 around the Hubble line for the GOODS SNe. We take $M_B = -19.09$, $\alpha = 0.14$, $\beta = 3.1$, $\delta = -0.07$, $\sigma_{\mathrm{sample}} = 0.11$ (Rubin et al., in prep) and find a distance modulus (no magnification or Malmquist bias correction) of \sndistmod. This is fully consistent with the value of \LCDMdistmod predicted from a flat $\Omega_m = 0.27$ $\Lambda$CDM universe. Figure \ref{fig:hubblediagram} shows the Hubble diagram of \citet{suzuki12} with \scpname and Primo \citep{rodney12} added. As SALT was updated from version 2-1 to 2-2 after this plot was made, we refit \scpname with the older SALT2-1 for the purposes of making this figure. The change in distance modulus is \DistChange magnitudes between the two versions.

The quality of these results at this extremely high redshift sets a new standard. Most SNe at $z > 1.5$ have incomplete or not cosmologically useful lightcurves (SN 1997ff from \citet{riess01}, 2003ak from \citet{riess04}, Subaru-Discovered SNe from \citet{graur11}). Primo \citep{rodney12} has a lower-precision color measurement than \scpname, although its better $x_1$ measurement (by virtue of pre-maximum data) gives it a comparable distance modulus error. All of these previous SNe had no spectroscopic confirmation, or in the case of Primo, a host-contaminated spectrum providing inconclusive confirmation.

It has appeared likely that SNe at this redshift could be measured with sufficient color precision to allow a direct comparison to lower-redshift SNe. With this one SN, we now see a first example of this in Figure \ref{fig:alphabeta}, a plot with a baseline of almost ten billion years (the approximate look back time of this SN). The Hubble residual of \scpname is compatible with the $x_1$ and $c$ corrections derived at lower redshift (or a deviation from $\Lambda$CDM of the Hubble diagram cancels the change in the relations). This figure also shows that the fitted $x_1$ and $c$ of \scpname are well within the distribution of lower-redshift supernovae that could be found in this F850LP search at redshift \twodecimalz (black points).

While the spectrum, light-curve corrections, and distance modulus of \scpname so far indicate compatibility with $\Lambda$CDM and little evolution, this single SN by itself can only provide weak constraints. It does, however, begin to illustrate what can be accomplished if one adds a whole population of such well-measured SNe at the very-high-redshift end of the Hubble diagram. Building this sample can now be done much more efficiently since the \textit{HST} WFC3 greatly improved throughput makes these high S/N measurements easier, so this goal is now within reach.

\appendix
\section{Weight of a Spectrum with Nearest-Neighbor Correlations}\label{sec:matinverse}

Suppose we have a spectrum with nearest-neighbor correlation $\rho$ between wavelength elements. We can write the spectrum covariance matrix as
\begin{equation}
C = \sigma \cdot (I + A) \cdot \sigma
\end{equation}
where $\sigma_{ij} = \sigma_i \delta_{ij}$, $I$ is the identity matrix, and $A_{ij} = \rho[\delta(|i - j| -1)]$. We would like the total weight of the spectrum, the sum of $C^{-1}$. Writing
\begin{equation}
C^{-1} = \sigma^{-1} \cdot (I + A)^{-1} \cdot \sigma ^{-1} \;,
\end{equation} we can focus on the $(I + A)^{-1}$ term. We begin by expanding this inverse as

\begin{equation}
(I + A)^{-1} = I + \sum_{k = 1}^{\infty} (-1)^k A^k
\end{equation}
We can now exchange the order of the matrix sum and series expansion and consider the sum of each term. The sum of $I$ is $N$, while for very large matrices (so that we can ignore edge effects), the sum of $A^k$ is $N (2 \rho)^k$, which goes to zero if $|\rho| < 1/2$. The desired sum is then
\begin{equation}
\sum_{ij} (I + A)^{-1} = N \sum_{k=0}^{\infty} (-2 \rho)^k = N/(1 + 2 \rho)
\end{equation}
for $|\rho| < 1/2$ as referenced in Section \ref{sec:acsgrism}.

\section{Spectral Principal Component Analysis}\label{sec:spectralmodel}
As discussed in Section \ref{sec:typing}, we use a principal component analysis to allow comparisons between spectra with limited wavelength coverage and non-negligible noise, as well as to help establish the dimensionality of the parameter space, so that $\Delta \chi^2$ values can be converted into probabilities. We have opted to perform this principal component analysis in $\log(\mathrm{flux})$ so that color variations can be more accurately modeled. As the signal-to-noise of most spectra is inadequate to simply take the log of the fluxes, we construct the principal components using an iterative fit.

We model each spectrum as
\begin{equation}
a_0*c_0 (\lambda) \exp[a_1*c_1 (\lambda) + a_2*c_2 (\lambda)]
\end{equation}
where $a_0$ is the normalization, $c_0(\lambda)$ represents the mean, $a_1$ and $a_2$ are the projections onto the first and second components, and $c_1 (\lambda)$ and $c_2 (\lambda)$ are the first and second components.

We fit the mean and first component (0 and 1, above) and their projections first (with the second component fixed to zero). After convergence, we fit the mean and second component with the first component held fixed. This sequential procedure ensures that at every stage, the component we are fitting is the one that contributes the most variance remaining. We start versions of the fit with many randomly chosen initial values for the projections to ensure that we have a converged solution (the components are always initialized at zero). We exclude the models of \citet{lentz00} from training the components, but we do compute the projections to enable a quantitative comparison to \scpname.

We use an error floor to prevent extremely well-measured wavelength regions or spectra from dominating the analysis. The error floor required is that needed to obtain a $\chi^2$ per degree of freedom of 1 for the residuals from the model. For our two-component analysis, this is S/N 5 per $\Delta \lambda /\lambda$ of 0.01 (a spectrum with $\Delta \lambda /\lambda$ of 0.001 would therefore be limited to S/N 1.6 per resolution element).

There is some ambiguity about how many principal components to use. Increasing the number allows for a smaller error floor (as more and more of the variance is described by the principal components). It also allows for better discrimination between spectra (e.g., spectra that are similar in the first two principal components may be dissimilar in the third). However, increasing the number also increases the $\Delta \chi^2$ values required for a given level of statistical significance. Two principal components are all that is necessary to fit almost all spectra to within the accuracy that the spectrum \scpname has been measured; two are therefore used for the results of this paper.

As a test, we also compute the probability of \scpname being a Ia (see Section \ref{sec:typing}) using one component and three components. Our results are robust; we find 93\% confidence using one component, 92\% confidence using two, and 91\% confidence using three components. It is important to note that we chose to use two components before seeing any of these probabilities.

\clearpage
\acknowledgements{
We would like to thank Henry Ferguson of the Space Telescope Science Institute for ensuring fast turnaround for these time-critical observations. We would also like to thank Bahram Mobasher for providing photometric redshifts for the host galaxies of our candidates. The archival WFC3 data used to obtain the host redshift were taken under HST GO Program 11600, PI Benjamin Weiner. We would like to thank the SUSPECT archive for their part in assembling our collection of spectra. Finally we thank the anonymous referee, whose feedback greatly improved this manuscript.

Financial support for this work was provided by NASA through program GO-9727 from the Space Telescope Science Institute, which is operated by AURA, Inc., under NASA contract NAS 5-26555. This work was also partially supported by the Director, Office of Science, Department of Energy, under grant DE-AC02-05CH11231.}

{\it Facilities:} \facility{Hubble Space Telescope}.

\bibliographystyle{apj}
\bibliography{ms}

\end{document}

%% file: numbers.tex
\newcommand{\scpname}{SN SCP-0401\xspace}

\newcommand{\snStoNz}{9\xspace} 
\newcommand{\snStoNi}{2\xspace} 
\newcommand{\snfoundmagz}{25.2\xspace} 
\newcommand{\snfoundmagi}{26.5\xspace} 

\newcommand{\sntocore}{$0.8\arcsec$\xspace} 
\newcommand{\sntocorekpc}{7 kpc\xspace} 
\newcommand{\nextclosest}{$3.5\arcsec$\xspace} 
\newcommand{\nogallimit}{AB $\sim26.5$ F775W\xspace} 

\newcommand{\nicJzp}{23.757 AB (23.029 Vega = 0)\xspace} 
\newcommand{\nicHzp}{22.16 (Vega = 0)\xspace} 
\newcommand{\zZP}{23.909 (Vega = 0)\xspace} 
\newcommand{\iZP}{25.291 (Vega = 0)\xspace} 
\newcommand{\nicHCRNLerr}{$0.026\pm0.015$ mag/dex\xspace} 

\newcommand{\cormagerr}{0.15\xspace} 
\newcommand{\sndistmod}{45.57 $\pm$ 0.24 statistical, $\pm \sim 0.1$ systematic\xspace} 
\newcommand{\LCDMdistmod}{45.60\xspace} 
 
\newcommand{\specphase}{$2 \pm 3$\xspace} 
\newcommand{\DistChange}{0.01\xspace}

\newcommand{\twodecimalz}{1.71\xspace} 
\newcommand{\galaxyz}{1.713\xspace} 
\newcommand{\galaxyzerr}{0.007\xspace}

\newcommand{\spiralstellarmass}{$4 \times 10^{10} M_{\odot}$\xspace}
\newcommand{\spiralhalomass}{$1.4 \times 10^{12} M_{\odot}$\xspace}

%% file: auth.tex
\author{
D. Rubin\altaffilmark{2, 3},
R. A. Knop\altaffilmark{4},
E. Rykoff\altaffilmark{2, 5},
G. Aldering\altaffilmark{2},
R. Amanullah\altaffilmark{6},
K. Barbary\altaffilmark{2},
M. S. Burns\altaffilmark{7},
A. Conley\altaffilmark{8},
N. Connolly\altaffilmark{9},
S. Deustua\altaffilmark{10},
V. Fadeyev\altaffilmark{11},
H. K. Fakhouri\altaffilmark{2, 3},
A. S. Fruchter\altaffilmark{10},
R. A. Gibbons\altaffilmark{12},
G. Goldhaber\altaffilmark{2, 3, 20},
A. Goobar\altaffilmark{6, 13},
E. Y. Hsiao\altaffilmark{2, 14, 15},
X. Huang\altaffilmark{3},
M. Kowalski\altaffilmark{16},
C. Lidman\altaffilmark{17},
J. Meyers\altaffilmark{2, 3},
J. Nordin\altaffilmark{2, 14},
S. Perlmutter\altaffilmark{2, 3},
C. Saunders\altaffilmark{2, 3},
A. L. Spadafora\altaffilmark{2},
V. Stanishev\altaffilmark{18},
N. Suzuki\altaffilmark{2, 14},
L. Wang\altaffilmark{19}
\\(The Supernova Cosmology Project)}

\altaffiltext{2}{E.O. Lawrence Berkeley National Lab, 1 Cyclotron Rd., Berkeley, CA, 94720}
\altaffiltext{3}{Department of Physics, University of California Berkeley, Berkeley, CA 94720}
\altaffiltext{4}{Quest University Canada, Squamish, BC, Canada.}
\altaffiltext{5}{Kavli Institute for Particle Astrophysics and Cosmology, SLAC National Accelerator Laboratory, Menlo Park, CA 94025}
\altaffiltext{6}{The Oskar Klein Centre, Department of Physics, AlbaNova, Stockholm University, SE-106 91 Stockholm, Sweden}
\altaffiltext{7}{Colorado College, 14 East Cache La Poudre St., Colorado Springs, CO 80903}
\altaffiltext{8}{Center for Astrophysics and Space Astronomy, 389 UCB, University of Colorado, Boulder, CO 80309}
\altaffiltext{9}{Hamilton College Department of Physics, Clinton, NY 13323}
\altaffiltext{10}{Space Telescope Science Institute, 3700 San Martin Drive, Baltimore, MD 21218}
\altaffiltext{11}{Santa Cruz Institute for Particle Physics, University of California Santa Cruz, Santa Cruze, CA 94064 }
\altaffiltext{12}{Department of Physics and Astronomy, Vanderbilt University, Nashville, TN 37240, USA}
\altaffiltext{13}{Department of Physics, Stockholm University, Albanova University Center, SE-106 91, Stockholm, Sweden}
\altaffiltext{14}{Space Sciences Lab, 7 Gauss Way, Berkeley, CA 94720}
\altaffiltext{15}{Carnegie Observatories, Las Campanas Observatory, Casilla 601, La Serena, Chile}
\altaffiltext{16}{Physikalisches Institut Universit\"at Bonn, Germany}
\altaffiltext{17}{Australian Astronomical Observatory, PO Box 296, Epping, NSW 1710, Australia}
\altaffiltext{18}{CENTRA - Centro Multidisciplinar de Astrof\'isica, Instituto Superior T\'ecnico, Av. Rovisco Pais 1, 1049-001 Lisbon, Portugal}
\altaffiltext{19}{Department of Physics, Texas A \& M University, College Station, TX 77843, USA}
\altaffiltext{20}{Deceased}

%% file: uvtable.tex
\tabletypesize{\scriptsize}
\setlength{\tabcolsep}{0.02in}
\begin{deluxetable*}{llllll}
\tablecolumns{6}
\tablecaption{Comparison Spectrum Sources}
\tablehead{
\colhead{SN} & \colhead{Type}  & \colhead{Type Reference} & \colhead{Phase}& \colhead{Date of Maximum Reference} & \colhead{Source and Program ID}}
\startdata
1978G  &II&\citet{1978IAUC.3309....1W} & Discovery +5, +16 & \citet{1978IAUC.3309....1W} &  \textit{IUE} OD7AB \\
1979C  &II&\citet{1979IAUC.3348....1M}&  +6 to +16  & \citet{1981PASP...93...36D} &  \textit{IUE} CVBCW, ESATO, UKTOO, CVBCW   \\
1980K  &II&\citet{1980IAUC.3534....2K}&  $\sim 0$  & \citet{1982PASP...94..578B} &  \textit{IUE} VILSP, CVBCW, UKTOO  \\
1980N  &  Ia  &\citet{1980IAUC.3556....2B}&  -1 to +12  & \citet{1991AJ....102..208H} &  \textit{IUE} CVBCW, VILSP  \\
1981B  &  Ia  &\citet{1981IAUC.3584....1V}&  +2, +3  & \citet{1983ApJ...270..123B} &  \textit{IUE} VILSP, NP314  \\
1982B  &  Ia  &\citet{1982IAUC.3671....1S}& +2& \citet{1988A+A...202...15C} &  \textit{IUE} NP586  \\
1983G  &  Ia  &\citet{1983IAUC.3791....2W}&  +3, +6, +9  & \citet{1985PASP...97..229B} &  \textit{IUE} SNFRK, FE022  \\
1983N  &  1b  &\citet{1985BASI...13...68P}&  -13 to +13  & N. Panagia, in \citet{2002ApJ...566.1005B} &  \textit{IUE} FE022, FETOO, SNFRK, OD15K  \\
1985L  &II&\citet{1985IAUC.4080....1F}& +12& \citet{1989Afz....31...17K} &  \textit{IUE} HQTOO  \\
1987A  &II&\citet{1987IAUC.4317....1H}&  -16, 0, +14  & \citet{1988A+A...198L...9G} &  \textit{IUE} OD17Y  \\
1989B  &  Ia  &\citet{1989BAVRu..38...88K}&  -9, -10  & \citet{1990A+A...232...75P} &  \textit{IUE} STKRK  \\
1989M  &  Ia  &\citet{1989IAUC.4802....1K}&  0 to +13  & \citet{1991SvA....35..168K} &  \textit{IUE} LETOO, SNLRK, LE059  \\
1990M  &  Ia  &\citet{1990IAUC.5034....1S}&  -6, -3  & \citet{1991A+A...242L...9P} &  \textit{IUE} SNMRK  \\
1990N  &Ia&\citet{1990IAUC.5039....1M}&  -10 to +4  & \citet{1991ApJ...371L..23L} &  \textit{IUE} SNMRK  \\
1990W  &  Ic  & \citet{1990IAUC.5079....1D} & +4& \citet{1990IAUC.5079....1D} &  \textit{IUE} SNMRK  \\
1991T  &  Ia  &\citet{1991IAUC.5251....1H}&  +8, +10  & \citet{1992AJ....103.1632P} &  \textit{IUE} METOO, SNMRK  \\
1992A  &  Ia  &\citet{1992IAUC.5428....1L}&  -2 to +11  & \citet{1992IAUC.5432....2S} &  \textit{IUE} SNNRK and \textit{HST} FOS 4016  \\
1993J  &  IIb  & \citet{1995JApAS..16..317P} &  -11, -3  & \citet{1995JApAS..16..317P} &  \textit{IUE} SNORK and \textit{HST} FOS 4528  \\
1994I  &  Ic  &\citet{1994IAUC.5964....1F}& +10& Richmond, in \citet{1994PASJ...46L.187S} &  \textit{HST} FOS 5623  \\
1997ap  &  Ia  & \citet{perlmutter97} &-2& \citet{perlmutter97} &  Keck II, \citet{perlmutter97}  \\
1998S  &II&\citet{1998IAUC.6829....1L}& +4& \citet{2000A+AS..144..219L} &  \textit{HST} STIS 7434  \\
1999em  &II&\citet{1999IAUC.7296....2J}& +5& \citet{2001ApJ...558..615H} &  \textit{HST} STIS 8243  \\
2001eh  &  Ia  &\citet{2001IAUC.7714....4G}& +7& SALT2-2 fit to \citet{2009ApJ...700..331H} &  \textit{HST} STIS 9114  \\
2001ep  &  Ia  &\citet{2001IAUC.7731....3M}&  +9, +15  & SALT2-2 fit to \citet{2009ApJ...700..331H} &  \textit{HST} STIS 9114   \\
2001ig  &  IIb  &\citet{2001IAUC.7772....2P}& Discovery +4, +12  & \citet{2001IAUC.7772....1E} &  \textit{HST} STIS 9114  \\
2002ap  &  Ic  &\citet{2002aprm.conf..333K}& -5& \citet{2003PASP..115.1220F} &  \textit{HST} STIS 9114  \\
2005cf  &  Ia  &\citet{2005CBET..160....1M}&  -9 to +4  & SALT2-2 fit to \citet{2009ApJ...700..331H} &  \textit{Swift} UVOT, \citet{bufano09}  \\
2005cs  &II&\citet{2005CBET..174....1M}&  +9, +11  & \citet{2009MNRAS.394.2266P} &  \textit{Swift} UVOT, \citet{bufano09}  \\
2006jc  &  Ib  &\citet{2005cxo..prop.2292I}&0& \citet{2007ApJ...657L.105F} &  \textit{Swift} UVOT, \citet{bufano09}  \\
2010al  &II&\citet{2010ATel.2513....1K}&$< 0$& \citet{2010ATel.2513....1K} &  \textit{HST} STIS 11654  \\
2011dh  &  IIb  &\citet{2011ATel.3428....1S}& +6& \citet{2012PZ.....32....6T} &  \textit{HST} STIS 12540  \\
2011iv  &  Ia  &\citet{2011CBET.2940....1D}& +1&\citet{foley12}&  \textit{HST} STIS 12592, \citet{foley12}  \\
SNLS  &  Ia  & \citet{ellis08} &0& \citet{ellis08} &  \citet{ellis08}  \\
Ia Model  &  Ia  & \citet{lentz00} &  0 (Explosion +20) & \citet{lentz00} &  \citet{lentz00} \label{tab:spectra}
\enddata
\tablecomments{Sources of data for the principal component analysis, indicating the SN type, source, and phase (phase range for many collected spectra from the same SN). \textit{IUE} is the \textit{International Ultraviolet Explorer}, \textit{HST} FOS/ STIS are the \textit{Hubble Space Telescope} Faint Object Spectrograph and Space Telescope Imaging Spectrograph, and \textit{Swift} UVOT is the \textit{Swift} Ultraviolet/Optical Telescope. The \textit{IUE} spectra extend blueward of $\sim 3300$\ang rest frame, the \textit{HST}, \textit{Swift}, and Lentz  spectra cover the whole wavelength range, the spectrum of 1997ap covers redward of $\sim 2700$\ang rest frame, and the Ellis composite covers redward of $\sim 2800$\ang rest frame.}
\end{deluxetable*}
\tabletypesize{\footnotesize}
\setlength{\tabcolsep}{0.1in}

%% file: probabilities.tex
\begin{deluxetable*}{llr}
\tablecolumns{3}
\tablecaption{Probabilities of matching \scpname.}
\tablehead{
\colhead{Supernova} & \colhead{Type} & \colhead{Probability of Match}}
\startdata
SN1980N &  Ia &  0.969 \\
SN2001ep &  Ia &  0.968 \\
SN1981B &  Ia &  0.963 \\
SN1992A &  Ia &  0.924 \\
SN2011iv &  Ia &  0.886 \\
SN1990N &  Ia &  0.610 \\
\citet{lentz00} & Ia Model &  0.514 \\
SN1983N &  Ib &  0.489 \\
SN2001eh &  Ia &  0.420 \\
SN1989M &  Ia &  0.316 \\
SN1982B &  Ia &  0.244 \\
SN1990M &  Ia &  0.157 \\
SN1989B &  Ia &  0.139 \\
SN1991T &  Ia &  0.059
\enddata
\label{tab:prob}
\tablecomments{Probabilities of each supernova matching \scpname. The values are taken from the principal-component-like analysis described in Section \ref{sec:typing} and Appendix \ref{sec:spectralmodel}. Only probabilities greater than 0.05 are shown.}
\end{deluxetable*}

%% file: photometry.tex
\begin{deluxetable*}{rrccrrrr}
\tablecolumns{9}
\tablecaption{Photometry of \scpname.}
\tablehead{
\colhead{MJD} & \colhead{PID} & \colhead{Camera} &\colhead{Filter} & \colhead{Exposure (s)} & \colhead{Flux (DN/s)} & \colhead{Flux Error (DN/s)} & \colhead{Vega=0 Zeropoint}
}
\startdata
52600.72 & 9583 & ACS WFC & F775W & 1120.0 & $-$0.0426 & 0.0599 & 25.291 \\
52600.75 & 9583 & ACS WFC & F850LP & 2400.0 & $-$0.0206 & 0.0274 & 23.909 \\
52643.38 & 9583 & ACS WFC & F775W & 1000.0 & $-$0.0271 & 0.0669 & 25.291 \\
52643.43 & 9583 & ACS WFC & F850LP & 2120.0 & 0.0180 & 0.0293 & 23.909 \\
52691.46 & 9583 & ACS WFC & F775W & 960.0 & $-$0.0422 & 0.0655 & 25.291 \\
52691.52 & 9583 & ACS WFC & F850LP & 2060.0 & 0.0014 & 0.0288 & 23.909 \\
52734.16 & 9583 & ACS WFC & F775W & 960.0 & 0.1358 & 0.0599 & 25.291 \\
52734.22 & 9583 & ACS WFC & F850LP & 2000.0 & $-$0.0128 & 0.0255 & 23.909 \\
52782.70 & 9583 & ACS WFC & F775W & 960.0 & 0.0864 & 0.0573 & 25.291 \\
52782.78 & 9583 & ACS WFC & F850LP & 2080.0 & $-$0.0048 & 0.0259 & 23.909 \\
53098.41 & 9727 & ACS WFC & F850LP & 1600.0 & 0.3249 & 0.0371 & 23.909 \\
53098.43 & 9727 & ACS WFC & F775W & 400.0 & 0.3328 & 0.1369 & 25.291 \\
53107.15 & 9727 & ACS WFC & F850LP & 4564.0 & 0.2213 & 0.0237 & 23.909 \\
53111.21 & 9727 & NICMOS 2 & F110W & 2687.9 & 0.2427 & 0.0144 & 23.029 \\
53111.31 & 9727 & NICMOS 2 & F160W & 5375.7 & 0.2011 & 0.0125 & 22.160 \\
53121.57 & 9727 & ACS WFC & F850LP & 4384.0 & 0.1311 & 0.0210 & 23.909 \\
53130.83 & 9727 & NICMOS 2 & F160W & 5375.7 & 0.1387 & 0.0120 & 22.160 \\
53136.86 & 9727 & NICMOS 2 & F110W & 2687.9 & 0.1205 & 0.0154 & 23.029 \\
53142.59 & 9727 & NICMOS 2 & F160W & 8063.6 & 0.1158 & 0.0103 & 22.160 \\
53145.98 & 9728 & ACS WFC & F775W & 400.0 & $-$0.0623 & 0.1536 & 25.291 \\
53146.01 & 9728 & ACS WFC & F850LP & 1600.0 & $-$0.0287 & 0.0361 & 23.909 \\
53148.41 & 9727 & NICMOS 2 & F110W & 8063.6 & 0.0811 & 0.0113 & 23.029 \\
53196.34 & 9727 & ACS WFC & F850LP & 1600.0 & 0.0536 & 0.0392 & 23.909 \\
53196.37 & 9727 & ACS WFC & F775W & 400.0 & 0.1695 & 0.1701 & 25.291 \\
53244.51 & 9728 & ACS WFC & F775W & 400.0 & $-$0.1195 & 0.1573 & 25.291 \\
53244.54 & 9728 & ACS WFC & F850LP & 1600.0 & 0.0003 & 0.0345 & 23.909 \\
53284.82 & 10339 & ACS WFC & F775W & 375.0 & $-$0.1097 & 0.1749 & 25.291 \\
53284.84 & 10339 & ACS WFC & F850LP & 1400.0 & $-$0.0009 & 0.0580 & 23.909 \\
53333.94 & 10339 & ACS WFC & F775W & 400.0 & $-$0.1821 & 0.1886 & 25.291 \\
53333.98 & 10339 & ACS WFC & F850LP & 1540.0 & $-$0.0756 & 0.0471 & 23.909 \\
53377.72 & 10339 & ACS WFC & F775W & 355.0 & $-$0.2258 & 0.2653 & 25.291 \\
53377.74 & 10339 & ACS WFC & F850LP & 1520.0 & $-$0.0214 & 0.0488 & 23.909 \\
53427.73 & 10339 & ACS WFC & F775W & 375.0 & 0.0921 & 0.2744 & 25.291 \\
53427.76 & 10339 & ACS WFC & F850LP & 1540.0 & 0.0346 & 0.0429 & 23.909 \\
53473.53 & 10339 & ACS WFC & F775W & 425.0 & $-$0.0096 & 0.1364 & 25.291 \\
53473.55 & 10339 & ACS WFC & F850LP & 1700.0 & 0.0359 & 0.0331 & 23.909
\enddata
\label{tab:phot}
\tablecomments{Due to the uncertainty on the galaxy models, the NICMOS F110W statistical errors share an off-diagonal covariance of 3.46e-5 DN/s$^2$, while the F160W errors share a separate off-diagonal covariance of 1.97e-5 DN/s$^2$. The ACS statistical errors are diagonal.}
\end{deluxetable*}